\documentclass[prd,preprint,superscriptaddress,amsmath,amssymb,nofootinbib]{revtex4}
\usepackage{graphicx}
\usepackage{dcolumn}
\usepackage{bm}
\usepackage{amssymb}
\usepackage{amsmath}
\usepackage{epsfig}    
\usepackage{color}
\usepackage{slashed}
\usepackage{hhline}
\usepackage{color}

\def\be{\begin{equation}}
\def\ee{\end{equation}}
\newcommand{\bea}{\begin{eqnarray}}
\newcommand{\eea}{\end{eqnarray}}
\newcommand{\nn}{\nonumber}

\begin{document}

{\begin{flushright}{APCTP Pre2020-021}
\end{flushright}}

\title{Lepton sector in modular $A_4$ and gauged $U(1)_R$ symmetry} 
\author{Keiko I. Nagao}
\email{nagao@dap.ous.ac.jp}
\affiliation{Okayama University of Science, Faculty of Science,  Department of Applied Physics, Ridaicho 1-1, Okayama, 700-0005, Japan}
\author{Hiroshi Okada}
\email{hiroshi.okada@apctp.org}
\affiliation{Asia Pacific Center for Theoretical Physics (APCTP) - Headquarters San 31, Hyoja-dong,
Nam-gu, Pohang 790-784, Korea}
\affiliation{Department of Physics, Pohang University of Science and Technology, Pohang 37673, Republic of Korea}

\date{\today}

\begin{abstract}
 We propose a lepton model under modular $A_4$ and gauged $U(1)_R$ symmetries, in which the neutrino masses are induced at one-loop level. 
 Thanks to the modular $A_4$ symmetry, we have several predictions on the lepton sector, especially, on fixed points of $\tau=i,\omega\equiv e^{2\pi i/3},i\times \infty$ each of which has remnant symmetry; $Z_2$ for $\tau=i$ and $Z_3$ for $\tau=\omega, i\infty$. These points are favored by a string theory and phenomenologically interesting. 
  \end{abstract}
\maketitle

\section{Introduction}
Right-handed gauged symmetry $U(1)_R$ is one of the promising candidates for physics beyond the standard model. It naturally accommodates the three right-handed fermions in order to cancel chiral anomalies, and also it and the other similar models such as gauged $U(1)_{B-L}$ symmetry can be discriminated by measuring, e.g., forward-backward asymmetry via each of the extra gauge bosons at International Linear Collider~\cite{Baer:2013cma}.

Recently, attractive flavor symmetries are also proposed by papers~\cite{Feruglio:2017spp,
deAdelhartToorop:2011re}, in which they applied modular motivated non-Abelian discrete flavor
symmetries to quark and lepton sectors.
One remarkable advantage is that any dimensionless couplings can also be transformed as non-trivial representations under
those symmetries. Therefore, we do not need so many scalars to find a predictive mass matrix.
Along this line of the idea,  a vast literature has recently arisen in references, {\it e.g.},  $A_4$~\cite{Feruglio:2017spp, Criado:2018thu, Kobayashi:2018scp, Okada:2018yrn, Nomura:2019jxj, Okada:2019uoy, deAnda:2018ecu, Novichkov:2018yse, Nomura:2019yft, Okada:2019mjf,Ding:2019zxk, Nomura:2019lnr,Kobayashi:2019xvz,Asaka:2019vev,Zhang:2019ngf, Gui-JunDing:2019wap,Kobayashi:2019gtp,Nomura:2019xsb, Wang:2019xbo,Okada:2020dmb,Okada:2020rjb, Behera:2020lpd, Behera:2020sfe, Nomura:2020opk, Nomura:2020cog, Asaka:2020tmo, Okada:2020ukr},
$S_3$ \cite{Kobayashi:2018vbk, Kobayashi:2018wkl, Kobayashi:2019rzp, Okada:2019xqk, Mishra:2020gxg}, $S_4$ \cite{Penedo:2018nmg, Novichkov:2018ovf, Kobayashi:2019mna,King:2019vhv,Okada:2019lzv,Criado:2019tzk,Wang:2019ovr}, $A_5$ \cite{Novichkov:2018nkm, Ding:2019xna,Criado:2019tzk}, larger groups~\cite{Baur:2019kwi}, multiple modular symmetries~\cite{deMedeirosVarzielas:2019cyj}, and double covering of $A_4$~\cite{Liu:2019khw, Chen:2020udk}, $S_4$~\cite{Novichkov:2020eep, Liu:2020akv}, and the other types of groups \cite{Kikuchi:2020nxn} in which  masses, mixing, and CP phases for quark and/or lepton are predicted.~\footnote{Some reviews are useful to understand the non-Abelian group and its applications to flavor structure~\cite{Altarelli:2010gt, Ishimori:2010au, Ishimori:2012zz, Hernandez:2012ra, King:2013eh, King:2014nza, King:2017guk, Petcov:2017ggy}.}
Furthermore, a systematic approach to understand the origin of CP transformations has been discussed in Ref.~\cite{Baur:2019iai}, 
and CP violation in models with modular symmetry is discussed in Ref.~\cite{Kobayashi:2019uyt,Novichkov:2019sqv}, 
and a possible correction from K\"ahler potential is discussed in Ref.~\cite{Chen:2019ewa}.
Systematic analysis of the fixed points(stabilizers) has been discussed in ref.~\cite{deMedeirosVarzielas:2020kji}.


In this paper, we combine these features of the gauged $U(1)_R$ based on our recent paper~\cite{Nagao:2020azf} and a modular $A_4$ symmetry, in which we construct lepton Yukawa Lagrangian while the neutrino mass  is generated by non-trivial Yukawa couplings at one-loop level.
Due to the $A_4$ nature, the charged-lepton  mass matrix is diagonal at the flavor eigenstate.
And we find specific Yukawa and mass matrices that lead to several predictions when we focus on three fixed points $\tau=i,\omega,i \infty$, where $\omega\equiv e^{2\pi i/3}$. These points retain remnant symmetries even after the breaking of flavor symmetry; $Z_2$ for $\tau=i$ and $Z_3$ for $\tau=\omega, i\infty$, which are favored by a string theory~\cite{Kobayashi:2020uaj}.

This letter is organized as follows.
In Sec. II, {we review our model, and formulate valid Higgs sector and lepton sector including heavier fermions, lepton flavor violations(LFVs), and muon anomalous magnetic moment.
In Sec. III, we have numerical analysis and show several predictions in each case of $\tau=i,\omega,i\infty$.
Finally we devote the summary to our results and the conclusion.}

\section{Model setup and Constraints}
\begin{table}[t!]
\begin{tabular}{|c||c|c|c|c|c||c|c|c|}\hline\hline  
& 
~$L_{L_1},L_{L_2},L_{L_3}$~& ~$e_{R_1},e_{R_2},e_{R_3}$~& ~$N_{R_a}$~& ~$S_{L_a}$~& ~$\bar L'_a$~& ~$H$~& ~$\varphi$~& ~$\chi$~\\\hline\hline 
$SU(3)_C$ 
& $\bm{1}$ & $\bm{1}$ & $\bm{1}$ & $\bm{1}$ & $\bm{1}$ & $\bm{1}$ & $\bm{1}$ & $\bm{1}$   \\\hline 
$SU(2)_L$ 
& $\bm{2}$  & $\bm{1}$  & $\bm{1}$  & $\bm{1}$ & $\bm{2}$   & $\bm{2}$ & $\bm{2}$   & $\bm{1}$     \\\hline 
$U(1)_Y$  
&  $-\frac12$  & $-1$ & $0$  & $0$  & $\frac12$ & $\frac12$  & $\frac12$& $0$  \\\hline
$U(1)_{R}$   
& $0$  & $-x$  & $x$  & $0$  & $x$ & $x$  & $2x$  & $x$\\\hline
$A_{4}$   
& ${\bf 1,1',1''}$  & ${\bf 1,1',1''}$  & $\bf3$  & $\bf3$ & $\bf3$ & $\bf1$   & $\bf1$ & $\bf1$\\\hline
$-k$   
& $0$  & $0$  & $-1$  & $-1$  & $-1$ & $0$  & $0$   & $-5$ \\\hline
\end{tabular}
\caption{ 
Charge assignments of the our fields
under $SU(3)_C\times SU(2)_L\times U(1)_Y\times U(1)_{R}$, where $SU(2)_L$ doublet quarks $Q_L$ and singlets $u_R, d_R$ in the SM have $0$, $x,-x$ under $U(1)_R$ symmetry, respectively,  in order to cancel chiral anomalies, and the upper index $a$ is the number of family that runs over 1-3.}
\label{tab:1}
\end{table}

In this section we depict our model.
In the fermion sector, we add three families of isospin singlet right-handed fermions $N_R$ with $x$ charge under the $U(1)_R$ gauge symmetry, where it is triplet under the $A_4$ symmetry and $-1$ under the modular weight $-k$. The detailed feature of modular $A_4$ described in the following is found in Appendix. 
This $x$ charge is required by chiral anomaly cancelations among fermions; $[U(1)_Y]^2U(1)_R$, $[U(1)_R]^2U(1)_Y$, $[U(1)_R]^3$, $U(1)_R$~\cite{Nomura:2016emz}. Notice here that quark sector in the SM also has to be nonzero charge of $U(1)_R$ symmetry;
$SU(2)_L$ doublet quarks $Q_L$ and singlets $u_R, d_R$ respectively have $0$, $x,-x$ under $U(1)_R$ symmetry in order to cancel the anomalies~\cite{Nagao:2020azf}.
Also three families of isospin singlet left-handed fermions $S_L$ and isospin doublet vector-like fermions $\bar L'_L$ are introduced 
in order to explain the neutrino mass matrix at one-loop level. Since $S_L$ has zero charge under $U(1)_R$, it does not affect the anomaly cancellations. Field $\bar L'_L$ also does not contribute to the anomaly cancellations because it is vector-like fermions. 
Here, both of $S_L$ and $\bar L'_L$ have the same charge assignments as the $N_R$'s charges under the $A_4$ symmetry and modular weight $-k$. 
In the boson sector, we accommodate the SM Higgs $H$ and two isospin singlet bosons $\varphi$ and $\chi$, where each of them has $U(1)_R$ charge of $x,2x,x$, respectively.
All the bosons are trivial singlet under the $A_4$ and only $\chi$ has nonzero modular weight $-5$.
We denote each of the bosons as follows here; 
 $ H \equiv [h^+, (v_{H}+h+iz)/\sqrt2]^T$, $\varphi \equiv (v' +r + i z')/\sqrt2$, and $\chi \equiv (\chi_R+i\chi_I)/\sqrt2$. Then, $h^+$, $z$  and $z'$  respectively give the nonzero masses for $W^+$, $Z$ and $Z'$ gauge boson after the spontaneously symmetry breaking, respectively.
 Field $\chi$ plays a role in generating the neutrino mass matrix at one-loop level.
All the field contents and their assignments are summarized in Table~\ref{tab:1}.
\footnote{Notice here that another realization is to assign $\chi$ to $-k=-3$ instead of $-5$.
However, in this case, we cannot satisfy one of the three observed mixings in PMNS matrix.
For example, we find the result of $\sin\theta_{13}^2\approx 0,0.3,1$ that are far from the experimental result of $\sim0.02$.} 

Under these symmetries, 
the valid Higgs potential is given by 
\begin{align}
{\it V}&= {\it V}_{trivial}+ \frac{\mu_1}2 (Y^{(10)}_1)^*\varphi^* \chi^2 
+{\rm h.c.},
\label{Eq:pot}
\end{align}
where we abbreviate the trivial quadratic and quartic terms that are proportional to $\phi^\dag \phi$, being $\phi=(H,\varphi,\chi)$.
Since $Y^{(10)}_1$ is a complex value, the real part of $\chi$ mixes with the imaginary one of $\chi$.
Thus the mass matrix is found as
\begin{align}
M^2_{inert}=\left[
\begin{array}{cc}
m^2+{\rm Re}[\delta m^2] &  -{\rm Im}[\delta m^2] \\ 
 -{\rm Im}[\delta m^2] &  m^2-{\rm Re}[\delta m^2]\\ 
\end{array}\right].\label{eq.minert}
\end{align}
Here, $m^2$ is constructed by trivial terms in ${\it V}_{trivial}$, while $\delta m^2$ is defined by $(Y^{(10)}_1)^* \mu_1 v_\varphi/\sqrt2$.
Then, $M^2_{inert}$ is diagonalized by an orthogonal matrix $O$ as ${\rm diag.}(m_1^2,m_2^2)=O M^2_{inert} O^T$. 
We denote the mass eigenstates as $H_1$ and $H_2$, and their masses are $m_1$ and $m_2$, respectively.
Even though each of $m_{1,2}$ is complicated form, we have simple relations in terms of components of $M^2_{inert}$ as follows:
\begin{align}
m^2_1+m^2_2 =2 m^2,\quad -m^2_1+m^2_2=2 |\delta m^2|=|Y^{(10)}_1| |\mu_1| v_\varphi/\sqrt2,\label{eq:pot-cond}
\end{align}
where $v_\varphi$ is supposed to be a real value.
If we assume to be $m^2>>\delta m^2$, we find $m^2_1=m^2_2=m^2$.
In cases of $\tau=i$ and $\tau=\omega$,  one finds $Y^{(10)}_1=0$, which implies that  $m^2>>\delta m^2$ could naturally be realized.
While, in case of $\tau=i\times\infty$,   $\tau=\omega$, one finds $Y^{(10)}_1=1$. Therefore, one has to impose the free mass parameter $\mu_1$ to be small enough to realize  $m^2>>\delta m^2$.

The Yukawa Lagrangian in the charged-lepton sector is diagonal due to the $A_4$ flavor symmetry and given as follows:
\begin{align}
 \sum_{\ell=1,2,3} y_\ell\bar L_{L_\ell}  H_1 e_{R_\ell} + {\rm h.c.}.\label{Eq:cgdl}
\end{align}
After the spontaneously symmetry breakings, the mass of charged-lepton is found by $m_e\equiv y_ev_{H_1}/\sqrt2$, $m_\mu\equiv y_\mu v_{H_1}/\sqrt2$, and $m_\tau\equiv y_\tau v_{H_1}/\sqrt2$.

An allowed Yukawa Lagrangian in the neutrino sector is given by
\begin{align}
&Y^{(6)}_{\rm 3} \otimes \bar L'_R \otimes  L_L \otimes \chi 
+Y^{(6)}_{\rm 3'} \otimes \bar L'_R \otimes  L_L \otimes \chi + {\rm h.c.}\nn\\
=&
\alpha_1(y'_1 \bar L'_{R_1} +y'_2 \bar L'_{R_3} +y'_3 \bar L'_{R_2}) L_{L_1}\chi
+
\beta_1(y'_2 \bar L'_{R_2} +y'_1 \bar L'_{R_3} +y'_3 \bar L'_{R_1}) L_{L_2}\chi\nn\\
&+
\gamma_1(y'_3 \bar L'_{R_3} +y'_1 \bar L'_{R_2} +y'_2 \bar L'_{R_1}) L_{L_3}\chi
\nn\\
+&
\alpha_2(y''_1 \bar L'_{R_1} +y''_2 \bar L'_{R_3} +y''_3 \bar L'_{R_2}) L_{L_1}\chi
+
\beta_2(y''_2 \bar L'_{R_2} +y''_1 \bar L'_{R_3} +y''_3 \bar L'_{R_1}) L_{L_2}\chi\nn\\
&+
\gamma_2(y''_3 \bar L'_{R_3} +y''_1 \bar L'_{R_2} +y''_2 \bar L'_{R_1}) L_{L_3}\chi
+ {\rm h.c.}
.\label{Eq:neut1}
\end{align}
Here, we define $Y^{(6)}_{3}\equiv [y'_1,y'_2,y'_3]^T$, $Y^{(6)}_{3'}\equiv [y''_1,y''_2,y''_3]^T$, and $\alpha_{1,2},\beta_{1,2},\gamma_{1,2}$ are free parameters.
This term is important to connect the sector of neutrino and exotic fields.
This gives the following Yukawa coupling
\begin{align}
f =
\left[
\begin{array}{ccc}
y'_1 & y'_3 &  y'_2  \\ 
 y'_3 &  y'_2 &  y'_1 \\ 
 y'_2  &  y'_1 &  y'_3 \\ 
\end{array}\right]
\left[
\begin{array}{ccc}
\alpha_1 &0 &  0  \\ 
0 &  \beta_1 &  0 \\ 
0  & 0 & \gamma_1 \\ 
\end{array}\right]
+
\left[
\begin{array}{ccc}
y''_1 & y''_3 &  y''_2  \\ 
 y''_3 &  y''_2 &  y''_1 \\ 
 y''_2  &  y''_1 &  y''_3 \\ 
\end{array}\right]
\left[
\begin{array}{ccc}
\alpha_2 &0 &  0  \\ 
0 &  \beta_2 &  0 \\ 
0  & 0 & \gamma_2 \\ 
\end{array}\right]
.\end{align}
By applying phase redefinitions of $L'_R$, we can suppose $\alpha_{1},\beta_1,\gamma_1$ as real parameters, while $\alpha_2,\beta_2,\gamma_2$ as complex, without loss of generality.

An allowed Yukawa Lagrangian that induces a neutral Dirac mass matrix is given by
\begin{align}
&(Y^{(2)}_{\rm 3})^* \otimes \bar S_L \otimes  L'_R \otimes H_2 + {\rm h.c.}\nn\\
=&
\frac a3
\left[y^*_1 (2\bar S_{L_1}L'_{R_1} -\bar S_{L_2}L'_{R_3} -\bar S_{L_3}L'_{R_2})
+y^*_3 (2\bar S_{L_3}L'_{R_3} -\bar S_{L_1}L'_{R_2} -\bar S_{L_2}L'_{R_1})\right.\nn\\
&\left.+y^*_2 (2\bar S_{L_2}L'_{R_2} -\bar S_{L_1}L'_{R_3} -\bar S_{L_3}L'_{R_1})
 \right]H_2 \nn\\
&+
\frac b2
\left[y^*_1 (\bar S_{L_3}L'_{R_2} -\bar S_{L_2}L'_{R_3})
+y^*_3 (\bar S_{L_2}L'_{R_1} -\bar S_{L_1}L'_{R_2})+y^*_2 (\bar S_{L_1}L'_{R_3} -\bar S_{L_3}L'_{R_1})
\right]H_2
+ {\rm h.c.},\label{Eq:dirac1}
\end{align}
where we define $Y^{(2)}_{3}\equiv [y_1,y_2,y_3]^T$, and $a,b$ are free parameters.
It gives the mass matrix after the spontaneously symmetry breaking as
\begin{align}
 m' =\frac{v_{H_2}}{\sqrt2} \left(
\frac a3\left[
\begin{array}{ccc}
2y^*_1 & -y^*_3 &  -y^*_2  \\ 
 -y^*_3 &  2y^*_2 &  -y^*_1 \\ 
 -y^*_2  &  -y^*_1 & 2 y^*_3 \\ 
\end{array}\right]
+
\frac b2\left[
\begin{array}{ccc}
0 & -y^*_3 &  y^*_2  \\ 
 y^*_3 & 0 &  -y^*_1 \\ 
 -y^*_2  &  y^*_1 &0 \\ 
\end{array}\right]
\right).
\end{align}
By applying phase redefinitions of $S_L$, we can consider $a$ as a real parameter, while $b$ as a complex one.

Another Yukawa Lagrangian to get a neutral Dirac mass matrix is given by
\begin{align}
&M_{L'}\otimes \bar L'_{L_L} \otimes L'_R + {\rm h.c.}
=
M_{L'}\left[ \bar L'_{L_1}  L'_{R_1} + \bar L'_{L_2}  L'_{R_2} + \bar L'_{L_3}  L'_{R_3} \right]
+ {\rm h.c.},\label{Eq:dirac2}
\end{align}
where $M_{L'}$, which is real mass parameter, includes invariant factor $1/(i\bar\tau-i\tau)$.

The right-handed Majorana mass matrix is given by
\begin{align}
&Y^{(2)}_{\rm 3} \otimes \bar N_R^C \otimes N_R\otimes \varphi^* + {\rm h.c.}\nn\\
=&
\frac{a_N}3
\left[y_1 (2\bar N^C_{R_1} N_{R_1} -\bar N^C_{R_2} N_{R_3} -\bar N^C_{R_3} N_{R_2})
+y_3 (2\bar N^C_{R_3} N_{R_3} -\bar N^C_{R_1} N_{R_2} -\bar N^C_{R_2} N_{R_1})\right.\nn\\
&\left.+y_2 (2\bar N^C_{R_2} N_{R_2} -\bar N^C_{R_1} N_{R_3} -\bar N^C_{R_3} N_{R_1})
 \right]  \varphi^*+ {\rm h.c.},\label{Eq:dirac1}
\end{align}
where  $a_N$ is a free parameter.
After the spontaneously symmetry breaking of $\varphi$, the mass matrix is given by
\begin{align}
M_R =
\frac{a_N v_\varphi}{3\sqrt2}\left[
\begin{array}{ccc}
2y_1 & -y_3 &  -y_2  \\ 
 -y_3 &  2y_2 &  -y_1 \\ 
 -y_2  &  -y_1 & 2 y_3 \\ 
\end{array}\right]
.\end{align}
By applying phase redefinitions of $N_R$, we can suppose $a_N$ as a real parameter.

The left-handed Majorana mass matrix is given by
\begin{align}
&M_0[ (Y^{(2)}_{\rm 3})^* \otimes \bar S_L \otimes S_L^C ]+ {\rm h.c.}\nn\\
=&
\frac{M_0}3
\left[y_1 (2\bar S_{L_1} S^C_{L_1} -\bar S_{L_2} S^C_{L_3} -\bar S_{L_3} S^C_{L_2})
+y_3 (2\bar S_{L_3} S^C_{L_3} -\bar S_{L_1} S^C_{L_2} -\bar S_{L_2} S^C_{L_1})\right.\nn\\
&\left.+y_2 (2\bar S_{L_2} S^C_{L_2} -\bar S_{L_1} S^C_{L_3} -\bar S_{L_3} S^C_{L_1})
 \right] + {\rm h.c.},\label{Eq:dirac1}
\end{align}
where $M_0$ is a free parameter.
Then, the mass matrix is given by
\begin{align}
M_S =
\frac{M_0}{3}\left[
\begin{array}{ccc}
2y_1^* & -y_3^* &  -y_2^*  \\ 
 -y_3^* &  2y_2^* &  -y_1^* \\ 
 -y_2^* &  -y_1^* & 2 y_3^* \\ 
\end{array}\right].
\end{align}

The neutral fermion mass matrix with 9$\times$9 based on $[N'_R,N'^c_L,S^c_L]^T$ is finally given by
\begin{align}
M_N
&=
\left[\begin{array}{ccc}
0 &   M_{L'}^T& 0  \\ 
M_{L'} & 0 & m' \\ 
0  &m'^T & M_{S} \\ 
\end{array}\right],
\end{align}
Then $M_N$ is diagonalized by a unitary matrix $V_N$ as $D_N\equiv V_N^T M_N V_N$ and $N=V_N\psi$, where $D_N$ is mass eigenvalue and $\psi$ is mass eigenstate.

The active neutrino mass matrix is given by~\cite{Ma:2006km} 
\begin{align}
(m_{\nu})_{ij}
&=\sum_{a=1}^{9}\sum_{k,k'=1}^{3}
\frac{f^T_{ik} (V_N^*)_{ka}D_{N_a} (V_N^\dag)_{ak'}f_{k'j}}{2 (4\pi)^2}
\left[ \frac{m^2_1}{m^2_1-D^2_{N_a}} \ln \frac{m^2_1}{D^2_{N_a}}-
\frac{m^2_2}{m^2_2-D^2_{N_a}} \ln \frac{m^2_2}{D^2_{N_a}}  \right] ~\nn\\
&\approx|\mu_1| \sum_{a=1}^{9}\sum_{k,k'=1}^{3}
\frac{f^T_{ik} (V_N^*)_{ka}D_{N_a} (V_N^\dag)_{ak'}f_{k'j}}{\sqrt2 (4\pi)^2}
\frac{v_\varphi |Y^{(10)}_1|}{m^2-D^2_{N_a}}
\left[1- \frac{D^2_{N_a} }{m^2 - D^2_{N_a}} \ln \frac{m^2}{D^2_{N_a}}  \right],
\end{align}
where we assume $m^2\ll \delta m^2$ in the second equation, and
$m_\nu$ is diagonalzied by a unitary matrix $U_{\rm PMNS}$~\cite{Maki:1962mu}; $D_\nu=|\mu_1| \tilde D_\nu= U_{\rm PMNS}^T m_\nu U_{\rm PMNS}=|\mu_1| U_{\rm PMNS}^T \tilde m_\nu U_{\rm PMNS}$.
Then $|\mu_1|$ is determined by
\begin{align}
(\mathrm{NO}):\  |\mu_1|^2= \frac{|\Delta m_{\rm atm}^2|}{\tilde D_{\nu_3}^2-\tilde D_{\nu_1}^2},
\quad
(\mathrm{IO}):\  |\mu_1|^2= \frac{|\Delta m_{\rm atm}^2|}{\tilde D_{\nu_2}^2-\tilde D_{\nu_3}^2},
 \end{align}
where $\Delta m_{\rm atm}^2$ is atmospheric neutrino mass difference squares, and NO and IO represent the normal hierarchy and the inverted hierarchy cases. 
Subsequently, the solar mass different squares can be written in terms of $|\mu_1|$ as follows:
\begin{align}
\Delta m_{\rm sol}^2=  |\mu_1|^2 ({\tilde D_{\nu_2}^2-\tilde D_{\nu_1}^2}),
 \end{align}
 which can be compared to the observed value.
 %
In our model, one finds $U_{\mathrm{PMNS}}=V_\nu$ since the charged-lepton is diagonal basis, and 
it is parametrized by three mixing angle $\theta_{ij} (i,j=1,2,3; i < j)$, one CP violating Dirac phase $\delta_{CP}$,
and two Majorana phases $\{\alpha_{21}, \alpha_{32}\}$ as follows:
\begin{equation}
U_{\mathrm{PMNS}} = 
\begin{pmatrix} c_{12} c_{13} & s_{12} c_{13} & s_{13} e^{-i \delta_{CP}} \\ 
-s_{12} c_{23} - c_{12} s_{23} s_{13} e^{i \delta_{CP}} & c_{12} c_{23} - s_{12} s_{23} s_{13} e^{i \delta_{CP}} & s_{23} c_{13} \\
s_{12} s_{23} - c_{12} c_{23} s_{13} e^{i \delta_{CP}} & -c_{12} s_{23} - s_{12} c_{23} s_{13} e^{i \delta_{CP}} & c_{23} c_{13} 
\end{pmatrix}
\begin{pmatrix} 1 & 0 & 0 \\ 0 & e^{i \frac{\alpha_{21}}{2}} & 0 \\ 0 & 0 & e^{i \frac{\alpha_{31}}{2}} \end{pmatrix},
\end{equation}
where $c_{ij}$ and $s_{ij}$ stands for $\cos \theta_{ij}$ and $\sin \theta_{ij}$ respectively. 
Then, each of mixing is given in terms of the component of $U_{\mathrm{PMNS}}$ as follows:
\begin{align}
\sin^2\theta_{13}=|(U_{\mathrm{PMNS}})_{13}|^2,\quad 
\sin^2\theta_{23}=\frac{|(U_{\mathrm{PMNS}})_{23}|^2}{1-|(U_{\mathrm{PMNS}})_{13}|^2},\quad 
\sin^2\theta_{12}=\frac{|(U_{\mathrm{PMNS}})_{12}|^2}{1-|(U_{\mathrm{PMNS}})_{13}|^2}.
\end{align}
Also we compute the Jarlskog invariant, $\delta_{CP}$ derived from PMNS matrix elements $U_{\alpha i}$:
\begin{equation}
J_{CP} = \text{Im} [U_{e1} U_{\mu 2} U_{e 2}^* U_{\mu 1}^*] = s_{23} c_{23} s_{12} c_{12} s_{13} c^2_{13} \sin \delta_{CP},
\end{equation}
and the Majorana phases are also estimated in terms of other invariants $I_1$ and $I_2$:
\begin{equation}
I_1 = \text{Im}[U^*_{e1} U_{e2}] = c_{12} s_{12} c_{13}^2 \sin \left( \frac{\alpha_{21}}{2} \right), \
I_2 = \text{Im}[U^*_{e1} U_{e3}] = c_{12} s_{13} c_{13} \sin \left( \frac{\alpha_{31}}{2}  - \delta_{CP} \right).
\end{equation}
In addition, the effective mass for the neutrinoless double beta decay is given by
\begin{align}
\langle m_{ee}\rangle=|\mu_1||\tilde D_{\nu_1} \cos^2\theta_{12} \cos^2\theta_{13}+\tilde D_{\nu_2} \sin^2\theta_{12} \cos^2\theta_{13}e^{i\alpha_{21}}+\tilde D_{\nu_3} \sin^2\theta_{13}e^{i(\alpha_{31}-2\delta_{CP})}|,
\end{align}
where its observed value could be measured by KamLAND-Zen in future~\cite{KamLAND-Zen:2016pfg}. 
We will adopt the neutrino experimental data at 3$\sigma$ interval~\cite{Esteban:2018azc} as follows:
\begin{align}
&{\rm NO}: \Delta m^2_{\rm atm}=[2.431, 2.622]\times 10^{-3}\ {\rm eV}^2,\
\Delta m^2_{\rm sol}=[6.79, 8.01]\times 10^{-5}\ {\rm eV}^2,\\
&\sin^2\theta_{13}=[0.02044, 0.02437],\ 
\sin^2\theta_{23}=[0.428, 0.624],\ 
\sin^2\theta_{12}=[0.275, 0.350],\nn\\
&{\rm IO}: \Delta m^2_{\rm atm}=[2.413, 2.606]\times 10^{-3}\ {\rm eV}^2,\
\Delta m^2_{\rm sol}=[6.79, 8.01]\times 10^{-5}\ {\rm eV}^2,\\
&\sin^2\theta_{13}=[0.02067, 0.02461],\ 
\sin^2\theta_{23}=[0.433, 0.623],\ 
\sin^2\theta_{12}=[0.275, 0.350].\nn
\end{align}

\subsection{Lepton flavor violations and anomalous magnetic moment} \label{lfv-lu}
Lepton flavor-violating (LFV) processes arise from the following Lagrangian
\begin{align}
{\cal L}_Y
&=
\frac{f_{ai}}{\sqrt2} \bar E'_{R_a} \ell_{L_i} (c_\theta H_1 - s_\theta H_2) +
\frac{f_{ai}}{\sqrt2} \bar E'_{R_a} \ell_{L_i} (s_\theta H_1 + c_\theta H_2) + \mbox{h.c.},
\label{eq:lvs-g2}
\end{align}
where $c_\theta(s_\theta)$ is a mixing angle of $O$ to diagonalize the inert boson mass matrix in Eq.(\ref{eq.minert}).
Then the branching ratio is given by
\begin{align}
{\rm BR}(\ell_i\to\ell_j\gamma)
&=
\frac{12\pi^3\alpha_{\rm em} C_{ij}c_\theta^2s_\theta^2}{(4\pi)^4G_F^2}\left(1+\frac{m^2_{\ell_j}}{m^2_{\ell_i}}\right)
\left|\sum_{\alpha=1}^3  f^\dag_{j\alpha} f_{\alpha i}  [F(m_1,M_{L'})- F(m_2,M_{L'})]\right|^2 ~,\\
F(m_a,m_b)&=\frac{m_b^6 -6 m_b^4 m_a^2 + 3 m_b^2 m_a^4 +2 m_a^6+6 m_b^2 m_a^4\ln\left[\frac{m^2_b}{m^2_a}\right]}{12(m_b^2-m_a^2)^4},
\label{eq:damu1}
\end{align}
where the fine structure constant $\alpha_{\rm em} \simeq 1/128$, the Fermi constant $G_F \simeq 1.17\times 10^{-5}$ GeV$^{-2}$, and $(C_{21},C_{31},C_{32}) \simeq (1,0.1784,0.1736)$.
The current experimental upper bounds at 90\% confidence level (CL) are~\cite{TheMEG:2016wtm, Adam:2013mnn}
\begin{align}
{\rm BR}(\mu\to e\gamma) < 4.2\times10^{-13} ~,~
{\rm BR}(\tau\to e\gamma) < 3.3\times10^{-8} ~,~
{\rm BR}(\tau\to \mu\gamma) < 4.4\times10^{-8} ~.
\end{align}
Muon $g-2$ is positively found via the same interaction with LFVs and its form is given by
 \begin{align}
 \Delta a_\mu^{(1)} \approx 
 \frac{m_\mu^2}{(4\pi)^2}s_\theta c_\theta
  \sum_{\alpha=1}^{3} f^\dag_{j\alpha} f_{\alpha i}  \left[F(m_1,M_{L'})- F(m_2,M_{L'})\right],
\label{amu1L}
 \end{align}
 where the discrepancy of the muon $g-2$ between the experimental measurement and the SM prediction is given by~\cite{Hagiwara:2011af} 
\begin{align}
\Delta a_{\mu}=(26.1 \pm 8.0)\times 10^{-10}. 
\end{align}
Since the mass difference between $H_1$ and $H_2$ is assumed to be tiny, $F(m_1,M_{L'})- F(m_2,M_{L'})$ is close to be zero. Thus, we do not need to consider the constraints of LFVs so seriously, even though we cannot obtain the muon $g-2$ enough. Here, we neglect them in our numerical analysis.

\section{Numerical analysis \label{sec:NA}}
\subsection{Normal Hierarchy Case \label{subsec:NA_NH}}
\begin{figure}[htbp]
 \begin{minipage}{0.32\hsize}
  \begin{center}
   \includegraphics[width=52mm]{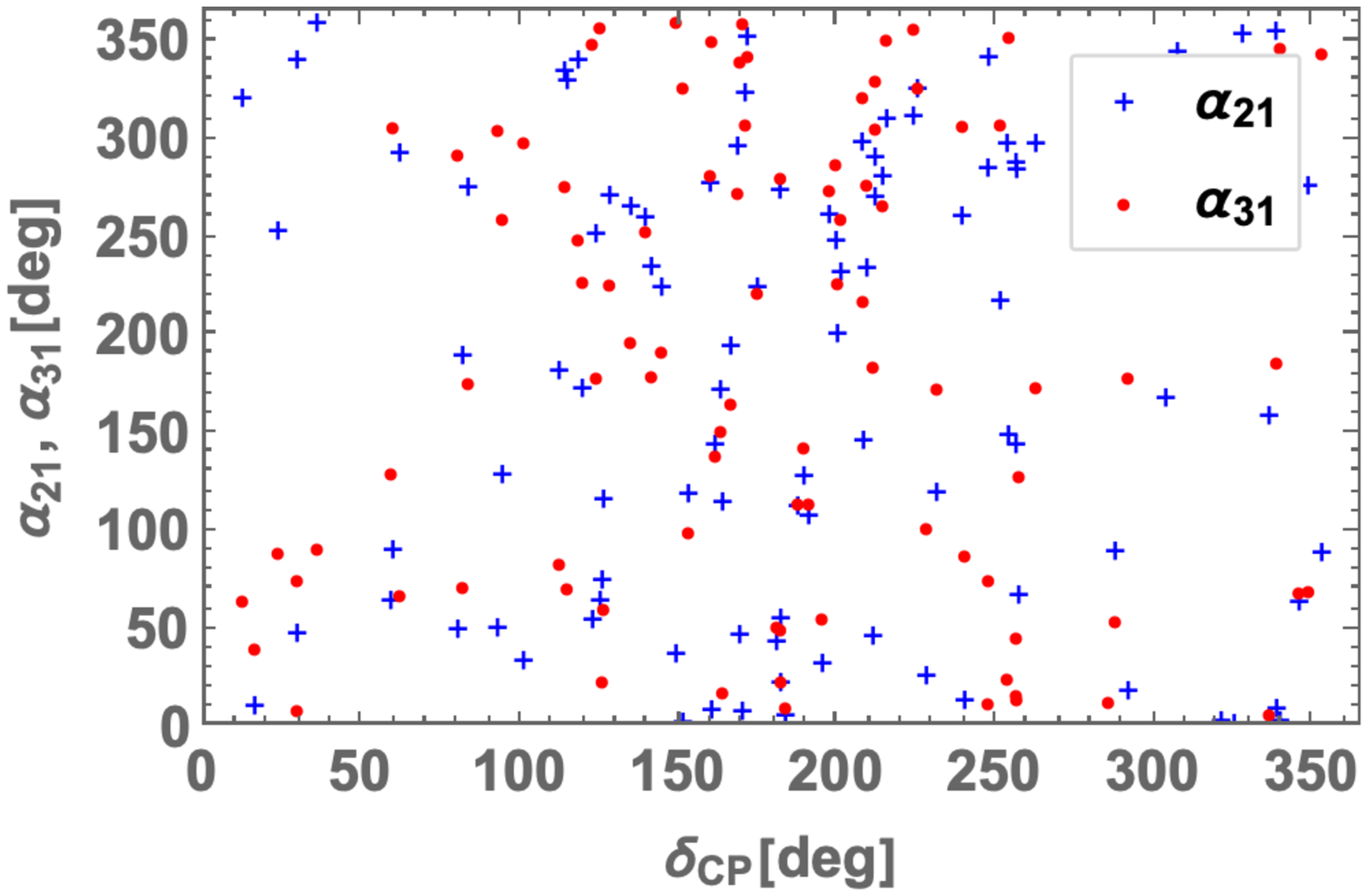}
  \end{center}
 \end{minipage}
 \begin{minipage}{0.32\hsize}
 \begin{center}
  \includegraphics[width=52mm]{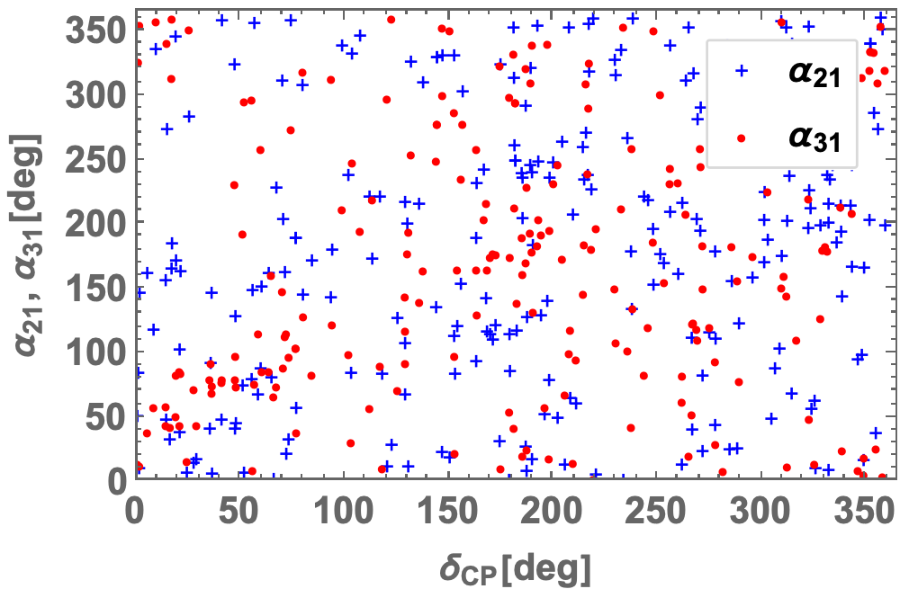}
 \end{center}
 \end{minipage}
 \begin{minipage}{0.32\hsize}
 \begin{center}
  \includegraphics[width=52mm]{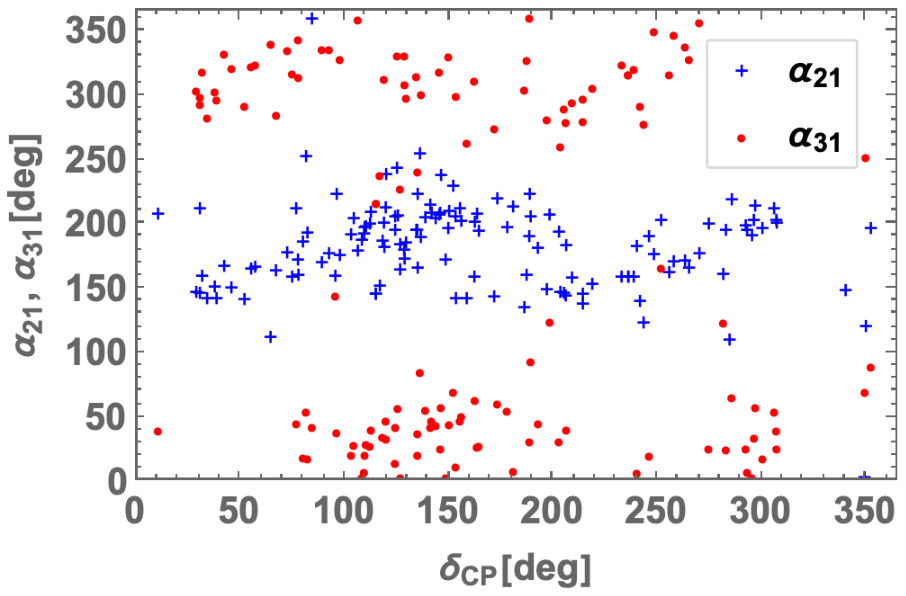}
 \end{center}
 \end{minipage}
   \caption{Plots of $\alpha_{12}$ and $\alpha_{31}$ for $\tau=\infty$ (left), $i$ (center) and $\omega$ (right), normal hierarchy case. Blue \textcolor{blue}{$+$} and Red \textcolor{red}{$\bullet$} represent $\alpha_{12}$ and $\alpha_{31}$ for corresponding $\delta_\mathrm{CP}$, respectively.}
   \label{fig:NHdelta}
\end{figure}
In Fig.~\ref{fig:NHdelta}, scatter plots of $\alpha_{12}$ and $\alpha_{31}$ are shown. No specific tendency of the distribution is seen for $\tau=\infty, i$ case. For $\tau=\omega$ case, on the other hand, we can see $100^\circ<\alpha_{21}<250^\circ$ and $\alpha_{31}<100^\circ, 250^\circ < \alpha_{31}$ are favored.

\begin{figure}[htbp] 
   \centering
   \includegraphics[width=80mm]{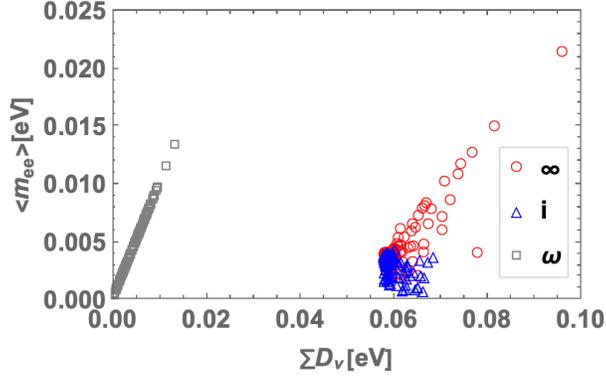} 
   \caption{Plot of $\sum D_v$--$\langle m_{ee}\rangle$ for $\tau=\infty$, $i$ and $\omega$, normal hierarchy case.}
   \label{fig:NHmass}
\end{figure}
In Fig.~\ref{fig:NHmass}, scatter plot of $\sum D_v$--$\langle m_{ee}\rangle$ is shown. For $\tau=\infty$ case, parameter region $0.05 < \sum D_v < 0.1$ eV and $\langle m_{ee}\rangle < 0.025$ eV is favored. For $\tau=i$ case, all the produced points in the numerical calculation is concentrated in $\sum D_v \sim 0.06$ eV and $\langle m_{ee}\rangle < 0.005$ eV. For $\tau=\omega$ case, we can see $\langle m_{ee}\rangle$ seems to be propotional to $D_v$.

 \begin{figure}[htbp] 
   \centering
   \includegraphics[width=80mm]{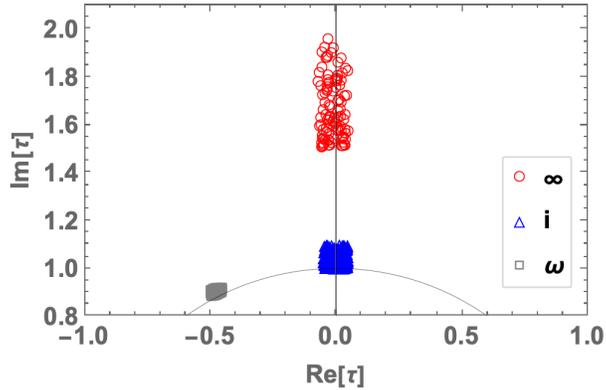} 
   \caption{Scatter plot of Re[$\tau$]--Im[$\tau$] for $\tau=\infty$, $i$ and $\omega$, normal hierarchy case.}
   \label{fig:NHReIm}
\end{figure}
In Fig.~\ref{fig:NHReIm}, scatter plot of Re[$\tau$]--Im[$\tau$] is shown. Points correspond to $\tau=\infty$ case distribute in Re[$\tau$]$\sim 0$, also $1.5 < $ Im[$\tau$] $< 2$. For $\tau=i$ and $\omega$ cases, the distribution is more dense than $\tau=\infty$ case. Especially, $\tau=\omega$ points are located not on the real axis, i.e., Re[$\tau$]$\sim -0.5$.

\subsection{Inverted Hierarchy Case}
\begin{figure}[htbp]
 \begin{minipage}{0.32\hsize}
  \begin{center}
   \includegraphics[width=52mm]{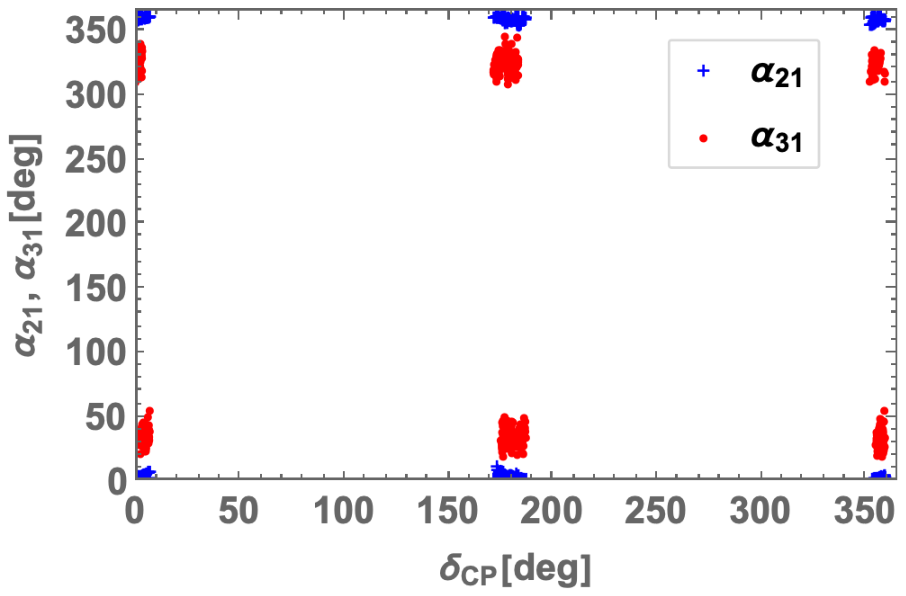}
  \end{center}
 \end{minipage}
 \begin{minipage}{0.32\hsize}
 \begin{center}
  \includegraphics[width=52mm]{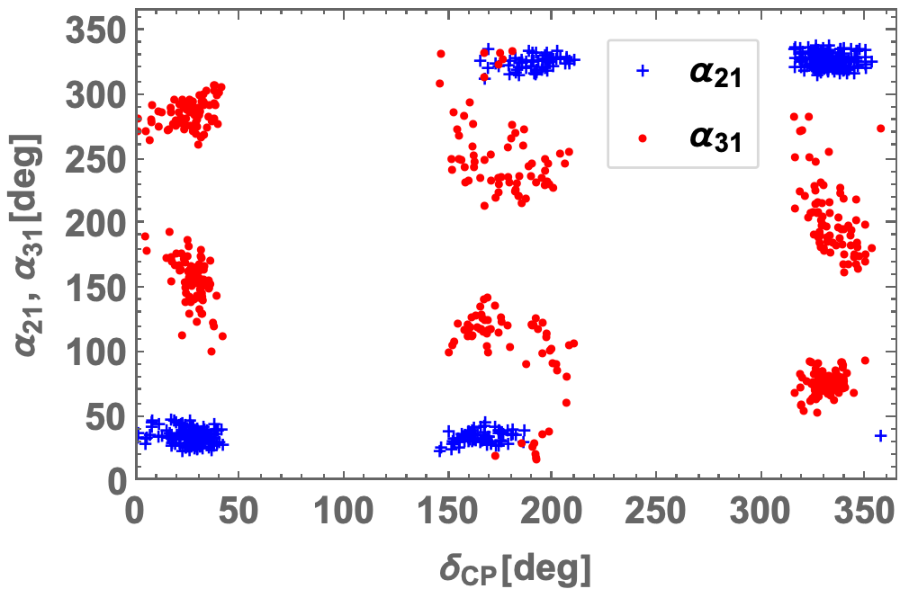}
 \end{center}
 \end{minipage}
 \begin{minipage}{0.32\hsize}
 \begin{center}
  \includegraphics[width=52mm]{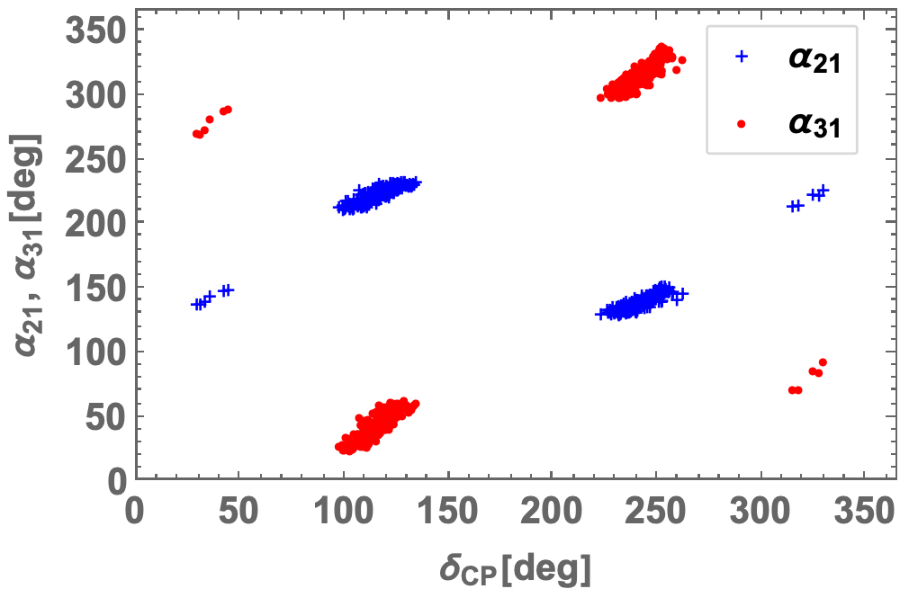}
 \end{center}
 \end{minipage}
   \caption{Legend is same as Fig.~\ref{fig:NHdelta} but for inverted hierarchy case.}
   \label{fig:IHdelta}
\end{figure}
In Fig.~\ref{fig:IHdelta}, scatter plots of $\alpha_{21}$ and $\alpha_{31}$ for inverted hierarchy are shown. Unlike the normal hierarchy case, several specific regions are favored even for $\tau=\infty, i$ cases. Especially, in $\tau=\infty$ case, all the points are densely gathered in region $\delta_\mathrm{CP}\sim 0^\circ (360^\circ), 180^\circ$ and $\alpha_{21}, \alpha_{31}\sim 0^\circ (360^\circ)$. 

\begin{figure}[htbp] 
   \centering
   \includegraphics[width=80mm]{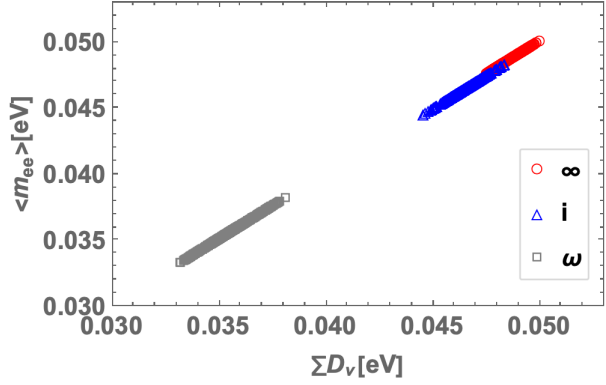} 
   \caption{Legend is same as Fig.~\ref{fig:NHmass} but for inverted hierarchy case.}
   \label{fig:IHmass}
\end{figure}
In Fig.~\ref{fig:IHmass}, scatter plot of $\sum D_v$--$\langle m_{ee}\rangle$ is shown. For all cases, $\langle m_{ee}\rangle$ seems to be propotional to $D_v$. Range $0.032 < \sum D_v < 0.038$, $0.044 < \sum D_v < 0.049$ and $0.047 < \sum D_v < 0.05$ on the proportional function are favored for $\tau=\infty, i, \omega$ cases, respectively.

 \begin{figure}[htbp] 
   \centering
   \includegraphics[width=80mm]{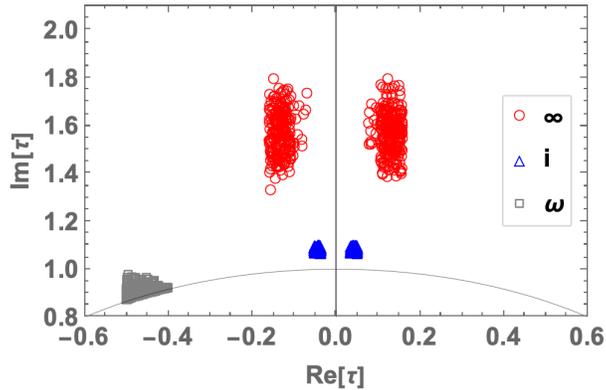} 
   \caption{Legend is same as Fig.~\ref{fig:NHReIm} but for inverted hierarchy case.}
   \label{fig:IHReIm}
\end{figure}
In Fig.~\ref{fig:IHReIm}, Scatter plot of Re[$\tau$]--Im[$\tau$] is shown. For $\tau=\infty, i$ cases, distributions are symmetric about the real axis. Points corresponds to $\tau=i$ case are converged on two narrow regions in small Im[$\tau$] region while points corresponds to $\tau=\infty$ widely spread in $1.3 <$ Im[$\tau$] $<1.8$ compared to $\tau=i$ case. Distribution of $\tau=\omega$ is concentrated on Re[$\tau$] $\sim - 0.5$ and Im[$\tau$] $\sim 0.9$, and not symmetric about the real axis.

\section{Summary and Conclusions}
 We have proposed a lepton model under the modular $A_4$ and gauged $U(1)_R$ symmetries, in which the neutrino masses are induced at one-loop level. 
Also we have several predictions on the lepton sector thanks to the modular $A_4$ symmetry, especially, on the fixed points of $\tau=i,\omega,i\times \infty$.
We especially point it out that we have found $100^\circ<\alpha_{21}<250^\circ$ and $\alpha_{31}<100^\circ, 250^\circ < \alpha_{31}$ are favored, and $0\lesssim \langle m_{ee}\rangle\lesssim0.02$ eV seems be proportional to $D_v$ for $\tau=\omega$ with NH.
And we have obtained more predictions for all the three cases with IH, as we have shown in the previous section.   
These cases would be verifiable and tested by future experiments soon.


\begin{acknowledgements}
KIN was also supported by  JSPS  Grant-in-Aid  for  ScientificResearch (A) 18H03699 and Wesco research grant.
This research was supported by an appointment to the JRG Program at the
APCTP through the Science and Technology Promotion Fund and Lottery Fund
of the Korean Government. This was also supported by the Korean Local
Governments - Gyeongsangbuk-do Province and Pohang City (H.O.).
H.O.~is sincerely grateful for the KIAS member. 
\end{acknowledgements}

\section*{Appendix}

Here, we show several properties of modular $A_4$ symmetry. 
In general, the modular group $\bar\Gamma$ is a group of linear fractional transformation
$\gamma$, acting on the modulus $\tau$ 
which belongs to the upper-half complex plane and transforms as
\begin{equation}\label{eq:tau-SL2Z}
\tau \longrightarrow \gamma\tau= \frac{a\tau + b}{c \tau + d}\ ,~~
{\rm where}~~ a,b,c,d \in \mathbb{Z}~~ {\rm and }~~ ad-bc=1, 
~~ {\rm Im} [\tau]>0 ~.
\end{equation}
This is isomorphic to  $PSL(2,\mathbb{Z})=SL(2,\mathbb{Z})/\{I,-I\}$ transformation.
Then modular transformation is generated by two transformations $S$ and $T$ defined by:
\begin{eqnarray}
S:\tau \longrightarrow -\frac{1}{\tau}\ , \qquad\qquad
T:\tau \longrightarrow \tau + 1\ ,
\end{eqnarray}
and they satisfy the following algebraic relations, 
\begin{equation}
S^2 =\mathbb{I}\ , \qquad (ST)^3 =\mathbb{I}\ .
\end{equation}
More concretely, we can fix the basis of $S$ and $T$ as follows:
  \begin{align}
S=\frac13
 \begin{pmatrix}
 -1 & 2 & 2  \\
 -2 & -1 & 2  \\
 2 & 2 & -1  \\
 \end{pmatrix} ,\quad 
 T= 
 \begin{pmatrix}
 1 & 0 & 0 \\
0 & \omega & 0  \\
0 & 0 & \omega^2  \\
 \end{pmatrix} ,
 \end{align}
where $\omega\equiv e^{2\pi i/3}$.

Here we introduce the series of groups $\Gamma(N)~ (N=1,2,3,\dots)$ which are defined by
 \begin{align}
 \begin{aligned}
 \Gamma(N)= \left \{ 
 \begin{pmatrix}
 a & b  \\
 c & d  
 \end{pmatrix} \in SL(2,\mathbb{Z})~ ,
 ~~
 \begin{pmatrix}
  a & b  \\
 c & d  
 \end{pmatrix} =
  \begin{pmatrix}
  1 & 0  \\
  0 & 1  
  \end{pmatrix} ~~({\rm mod}~N) \right \}
 \end{aligned},
 \end{align}
and we define $\bar\Gamma(2)\equiv \Gamma(2)/\{I,-I\}$ for $N=2$.
Since the element $-I$ does not belong to $\Gamma(N)$
  for $N>2$ case, we have $\bar\Gamma(N)= \Gamma(N)$,
  that are infinite normal subgroup of $\bar \Gamma$ known as principal congruence subgroups.
   We thus obtain finite modular groups as the quotient groups defined by
   $\Gamma_N\equiv \bar \Gamma/\bar \Gamma(N)$.
For these finite groups $\Gamma_N$, $T^N=\mathbb{I}$  is imposed, and
the groups $\Gamma_N$ with $N=2,3,4$ and $5$ are isomorphic to
$S_3$, $A_4$, $S_4$ and $A_5$, respectively \cite{deAdelhartToorop:2011re}.

Modular forms of level $N$ are 
holomorphic functions $f(\tau)$ which are transformed under the action of $\Gamma(N)$ given by
\begin{equation}
f(\gamma\tau)= (c\tau+d)^k f(\tau)~, ~~ \gamma \in \Gamma(N)~ ,
\end{equation}
where $k$ is the so-called as the  modular weight.

Under the modular transformation in Eq.(\ref{eq:tau-SL2Z}) in case of $A_4$ ($N=3$) modular group, a field $\phi^{(I)}$ is also transformed as 
\begin{equation}
\phi^{(I)} \to (c\tau+d)^{-k_I}\rho^{(I)}(\gamma)\phi^{(I)},
\end{equation}
where  $-k_I$ is the modular weight and $\rho^{(I)}(\gamma)$ denotes a unitary representation matrix of $\gamma\in\Gamma(2)$ ($A_4$ reperesantation).
Thus Lagrangian such as Yukawa terms can be invariant if sum of modular weight from fields and modular form in corresponding term is zero (also invariant under $A_4$ and gauge symmetry).

The kinetic terms and quadratic terms of scalar fields can be written by 
\begin{equation}
\sum_I\frac{|\partial_\mu\phi^{(I)}|^2}{(-i\tau+i\bar{\tau})^{k_I}} ~, \quad \sum_I\frac{|\phi^{(I)}|^2}{(-i\tau+i\bar{\tau})^{k_I}} ~,
\label{kinetic}
\end{equation}
which is invariant under the modular transformation and overall factor is eventually absorbed by a field redefinition consistently.
Therefore the Lagrangian associated with these terms should be invariant under the modular symmetry.

The basis of modular forms with weight 2, $Y^{(2)}_3 = (y_{1},y_{2},y_{3})$, transforming
as a triplet of $A_4$ is written in terms of Dedekind eta-function  $\eta(\tau)$ and its derivative \cite{Feruglio:2017spp}:
\begin{align} 
\label{eq:Y-A4}
y_{1}(\tau) &= \frac{i}{2\pi}\left( \frac{\eta'(\tau/3)}{\eta(\tau/3)}  +\frac{\eta'((\tau +1)/3)}{\eta((\tau+1)/3)}  
+\frac{\eta'((\tau +2)/3)}{\eta((\tau+2)/3)} - \frac{27\eta'(3\tau)}{\eta(3\tau)}  \right)\nn\\ 
&\simeq
1+12 q+36 q^2+12 q^3+\cdots,\\
y_{2}(\tau) &= \frac{-i}{\pi}\left( \frac{\eta'(\tau/3)}{\eta(\tau/3)}  +\omega^2\frac{\eta'((\tau +1)/3)}{\eta((\tau+1)/3)}  
+\omega \frac{\eta'((\tau +2)/3)}{\eta((\tau+2)/3)}  \right) , \label{eq:Yi} \nn\\ 
&\simeq
-6q^{1/3} (1+7 q+8 q^2+\cdots),\\
y_{3}(\tau) &= \frac{-i}{\pi}\left( \frac{\eta'(\tau/3)}{\eta(\tau/3)}  +\omega\frac{\eta'((\tau +1)/3)}{\eta((\tau+1)/3)}  
+\omega^2 \frac{\eta'((\tau +2)/3)}{\eta((\tau+2)/3)}  \right)\nn\\ 
&\simeq
-18q^{2/3} (1+2 q+5 q^2+\cdots),
\end{align}
where $q=e^{2\pi i \tau}$, and expansion form in terms of $q$ would sometimes be useful to have numerical analysis.

Then, we can construct the higher order of couplings $Y^{(4)}_1,Y^{(6)}_1,Y^{(10)}_1,  Y^{(6)}_3, Y^{(6)}_{3'}$ following the multiplication rules as follows:
\begin{align}
Y^{(4)}_1& = y^2_1+2 y_2 y_3, \ Y^{(6)}_1 = y^3_1+y^3_2 + y^3_3 -3 y_1y_2y_3, \
 Y^{(10)}_1 = Y^{(4)}_1 Y^{(6)}_1,\\
Y^{(6)}_3&\equiv (y'_1,y'_2,y'_3) = ( y^3_1+2y_1 y_2 y_3, y_1^2y_2+2 y^2_2 y_3, y^2_1 y_3+2 y^2_3 y_2),\\
Y^{(6)}_{3'}&\equiv (y''_1,y''_2,y''_3) = ( y^3_3+2y_1 y_2 y_3, y^2_3 y_1+2 y^2_1 y_2, y^2_3 y_2+2 y^2_2 y_1),
\end{align}
where  the above relations are constructed by the multiplication rules under $A_4$ as shown below:
\begin{align}
\begin{pmatrix}
a_1\\
a_2\\
a_3
\end{pmatrix}_{\bf 3}
\otimes 
\begin{pmatrix}
b_1\\
b_2\\
b_3
\end{pmatrix}_{\bf 3'}
&=\left (a_1b_1+a_2b_3+a_3b_2\right )_{\bf 1} 
\oplus \left (a_3b_3+a_1b_2+a_2b_1\right )_{{\bf 1}'} \nonumber \\
& \oplus \left (a_2b_2+a_1b_3+a_3b_1\right )_{{\bf 1}''} \nonumber \\
&\oplus \frac13
\begin{pmatrix}
2a_1b_1-a_2b_3-a_3b_2 \\
2a_3b_3-a_1b_2-a_2b_1 \\
2a_2b_2-a_1b_3-a_3b_1
\end{pmatrix}_{{\bf 3}}
\oplus \frac12
\begin{pmatrix}
a_2b_3-a_3b_2 \\
a_1b_2-a_2b_1 \\
a_3b_1-a_1b_3
\end{pmatrix}_{{\bf 3'}\  } \ , \nonumber \\
\nonumber \\
{\bf 1} \otimes {\bf 1} = {\bf 1} \ , \quad &
{\bf 1'} \otimes {\bf 1'} = {\bf 1''} \ , \quad
{\bf 1''} \otimes {\bf 1''} = {\bf 1'} \ , \quad
{\bf 1'} \otimes {\bf 1''} = {\bf 1} \ .
\end{align}

Finally, we show the features of fixed points of $\tau=i,\omega, i\times \infty$.\\
\begin{itemize}
\item In case of $\tau=i$, it is invariant under the transformation of $\tau\to -1/\tau$ that corresponds to $S$ transformation.
It implies that there is a remnant $Z_2$ symmetry and its element is given by $\{1,S\}$.
Then, the concrete value of $Y^{(2)}_3$ can be written down by~\cite{Okada:2020ukr}
\begin{align}
Y^{(2)}_3&\simeq1.0025(1,1-\sqrt3,-2+\sqrt3).
\end{align}
 \item In case of $\tau=\omega$, it is invariant under the transformation of $\tau\to -1/(1+\tau)$ that corresponds to $ST$ transformation.
It implies that there is a remnant $Z_3$ symmetry and its element is given by $\{1,ST,(ST)^2\}$.
Then, the concrete value of $Y^{(2)}_3$ can be written down by~\cite{Okada:2020ukr}
\begin{align}
Y^{(2)}_3&\simeq 0.9486(1,\omega, -\frac12\omega^2).
\end{align}
\item In case of $\tau=i\times\infty$, this corresponds to $T$ transformation.
It suggests that there is a remnant $Z_3$ symmetry and its element is given by $\{1,T,T^2\}$.
Then, the concrete value of $Y^{(2)}_3$ can be written down by~\cite{Okada:2020ukr}
\begin{align}
Y^{(2)}_3&\simeq (1,0,0).
\end{align}
\end{itemize}

\end{document}